\documentclass[12pt]{article}
\usepackage{epsfig}
\usepackage{amsmath}
\usepackage{hhline}
\usepackage{amssymb}
\usepackage{times}
\usepackage{cite}

\newlength{\dinwidth}
\newlength{\dinmargin}
\begin{document}  
\newcommand{\pom}{{I\!\!P}}
\newcommand{\reg}{{I\!\!R}}
\newcommand{\slowpi}{\pi_{\mathit{slow}}}
\newcommand{\fiidiii}{F_2^{D(3)}}
\newcommand{\fiidiiiarg}{\fiidiii\,(\beta,\,Q^2,\,x)}
\newcommand{\n}{1.19\pm 0.06 (stat.) \pm0.07 (syst.)}
\newcommand{\nz}{1.30\pm 0.08 (stat.)^{+0.08}_{-0.14} (syst.)}
\newcommand{\fiidiiiful}{F_2^{D(4)}\,(\beta,\,Q^2,\,x,\,t)}
\newcommand{\fiipom}{\tilde F_2^D}
\newcommand{\ALPHA}{1.10\pm0.03 (stat.) \pm0.04 (syst.)}
\newcommand{\ALPHAZ}{1.15\pm0.04 (stat.)^{+0.04}_{-0.07} (syst.)}
\newcommand{\fiipomarg}{\fiipom\,(\beta,\,Q^2)}
\newcommand{\pomflux}{f_{\pom / p}}
\newcommand{\nxpom}{1.19\pm 0.06 (stat.) \pm0.07 (syst.)}
\newcommand {\gapprox}
   {\raisebox{-0.7ex}{$\stackrel {\textstyle>}{\sim}$}}
\newcommand {\lapprox}
   {\raisebox{-0.7ex}{$\stackrel {\textstyle<}{\sim}$}}
\def\gsim{\,\lower.25ex\hbox{$\scriptstyle\sim$}\kern-1.30ex%
\raise 0.55ex\hbox{$\scriptstyle >$}\,}
\def\lsim{\,\lower.25ex\hbox{$\scriptstyle\sim$}\kern-1.30ex%
\raise 0.55ex\hbox{$\scriptstyle <$}\,}
\newcommand{\pomfluxarg}{f_{\pom / p}\,(x_\pom)}
\newcommand{\dsf}{\mbox{$F_2^{D(3)}$}}
\newcommand{\dsfva}{\mbox{$F_2^{D(3)}(\beta,Q^2,x_{I\!\!P})$}}
\newcommand{\dsfvb}{\mbox{$F_2^{D(3)}(\beta,Q^2,x)$}}
\newcommand{\dsfpom}{$F_2^{I\!\!P}$}
\newcommand{\gap}{\stackrel{>}{\sim}}
\newcommand{\lap}{\stackrel{<}{\sim}}
\newcommand{\fem}{$F_2^{em}$}
\newcommand{\tsnmp}{$\tilde{\sigma}_{NC}(e^{\mp})$}
\newcommand{\tsnm}{$\tilde{\sigma}_{NC}(e^-)$}
\newcommand{\tsnp}{$\tilde{\sigma}_{NC}(e^+)$}
\newcommand{\st}{$\star$}
\newcommand{\sst}{$\star \star$}
\newcommand{\ssst}{$\star \star \star$}
\newcommand{\sssst}{$\star \star \star \star$}
\newcommand{\tw}{\theta_W}
\newcommand{\sw}{\sin{\theta_W}}
\newcommand{\cw}{\cos{\theta_W}}
\newcommand{\sww}{\sin^2{\theta_W}}
\newcommand{\cww}{\cos^2{\theta_W}}
\newcommand{\trm}{m_{\perp}}
\newcommand{\trp}{p_{\perp}}
\newcommand{\trmm}{m_{\perp}^2}
\newcommand{\trpp}{p_{\perp}^2}
\newcommand{\alp}{\alpha_s}

\newcommand{\alps}{\alpha_s}
\newcommand{\sqrts}{$\sqrt{s}$}
\newcommand{\LO}{$O(\alpha_s^0)$}
\newcommand{\Oa}{$O(\alpha_s)$}
\newcommand{\Oaa}{$O(\alpha_s^2)$}
\newcommand{\PT}{p_{\perp}}
\newcommand{\JPSI}{J/\psi}
\newcommand{\sh}{\hat{s}}
\newcommand{\uh}{\hat{u}}
\newcommand{\MP}{m_{J/\psi}}
\newcommand{\PO}{I\!\!P}
\newcommand{\xbj}{x}
\newcommand{\xpom}{x_{\PO}}
\newcommand{\ttbs}{\char'134}
\newcommand{\xpomlo}{3\times10^{-4}}  
\newcommand{\xpomup}{0.05}  
\newcommand{\dgr}{^\circ}
\newcommand{\pbarnt}{\,\mbox{{\rm pb$^{-1}$}}}
\newcommand{\gev}{\,\mbox{GeV}}
\newcommand{\WBoson}{\mbox{$W$}}
\newcommand{\fbarn}{\,\mbox{{\rm fb}}}
\newcommand{\fbarnt}{\,\mbox{{\rm fb$^{-1}$}}}
%
%
\newcommand{\qsq}{\ensuremath{Q^2} }
\newcommand{\gevsq}{\ensuremath{\mathrm{GeV}^2} }
\newcommand{\et}{\ensuremath{E_t^*} }
\newcommand{\rap}{\ensuremath{\eta^*} }
\newcommand{\gp}{\ensuremath{\gamma^*}p }
\newcommand{\dsiget}{\ensuremath{{\rm d}\sigma_{ep}/{\rm d}E_t^*} }
\newcommand{\dsigrap}{\ensuremath{{\rm d}\sigma_{ep}/{\rm d}\eta^*} }
\def\Journal#1#2#3#4{{#1} {\bf #2} (#3) #4}
\def\NCA{\em Nuovo Cimento}
\def\NIM{\em Nucl. Instrum. Methods}
\def\NIMA{{\em Nucl. Instrum. Methods} {\bf A}}
\def\NPB{{\em Nucl. Phys.}   {\bf B}}
\def\PLB{{\em Phys. Lett.}   {\bf B}}
\def\PRL{\em Phys. Rev. Lett.}
\def\PRD{{\em Phys. Rev.}    {\bf D}}
\def\ZPC{{\em Z. Phys.}      {\bf C}}
\def\EJC{{\em Eur. Phys. J.} {\bf C}}
\def\CPC{\em Comp. Phys. Commun.}

\newcommand{\dps}{\displaystyle}

\begin{titlepage}

\noindent
DESY 01-178  \hfill  ISSN 0418-9833 \\
October 2001

\vspace{3cm}

\begin{center}
\begin{Large}

\boldmath

{\bf Measurement of Dijet Electroproduction \\
at Small Jet Separation}

\unboldmath

\vspace{2cm}

H1 Collaboration

\end{Large}
\end{center}

\vspace{2cm}

\begin{abstract}

\noindent Deep-inelastic scattering data in the range $150 < Q^2 <
35\,000\,$GeV$^2$ are used to investigate the minimum jet
separation necessary to allow accurate description of the rate of
dijet production using next-to-leading order perturbative QCD
calculations. The required jet separation is found to be small,
allowing about $1/3$ of DIS data to be classified as dijet, as
opposed to approximately $1/10$ with more typical jet analyses. A number of
precision measurements made using this dijet sample are well
described by the calculations. The data are also described
by the combination of leading order matrix elements and parton
showers, as implemented in the QCD based Monte Carlo model RAPGAP.

\end{abstract}

\vspace{1.5cm}

\centerline{\it To be submitted to the European Physical Journal }

\end{titlepage}

\clearpage

\newpage

\noindent
C.~Adloff$^{33}$,              
V.~Andreev$^{24}$,             
B.~Andrieu$^{27}$,             
T.~Anthonis$^{4}$,             
V.~Arkadov$^{35}$,             
A.~Astvatsatourov$^{35}$,      
A.~Babaev$^{23}$,              
J.~B\"ahr$^{35}$,              
P.~Baranov$^{24}$,             
E.~Barrelet$^{28}$
% ...

\bigskip{\it
\noindent
% ...
$ ^{2}$ III. Physikalisches Institut der RWTH, Aachen, Germany$^{ a}$ \\
% ...
$ ^{4}$ Inter-University Institute for High Energies ULB-VUB, Brussels;
          Universitaire Instelling Antwerpen, Wilrijk; Belgium$^{ c}$ \\
% ...
$ ^{23}$ Institute for Theoretical and Experimental Physics,
          Moscow, Russia$^{ l}$ \\
 $ ^{24}$ Lebedev Physical Institute, Moscow, Russia$^{ e,h}$ \\
% ...
$ ^{27}$ LPNHE, Ecole Polytechnique, IN2P3-CNRS, Palaiseau, France \\
 $ ^{28}$ LPNHE, Universit\'{e}s Paris VI and VII, IN2P3-CNRS,
          Paris, France \\
% ...
$ ^{33}$ Fachbereich Physik, Bergische Universit\"at Gesamthochschule
          Wuppertal, Wuppertal, Germany \\
% ...
$ ^{35}$ DESY, Zeuthen, Germany \\
% ...
$ ^a$ Supported by the Bundesministerium f\"ur Bildung und Forschung, FRG,
      under contract numbers 05 H1 1GUA /1, 05 H1 1PAA /1, 05 H1 1PAB /9,
      05 H1 1PEA /6, 05 H1 1VHA /7 and 05 H1 1VHB /5 \\
% ...
$ ^c$ Supported by FNRS-NFWO, IISN-IIKW \\
% ...
$ ^e$ Supported by the Deutsche Forschungsgemeinschaft \\
% ...
$ ^h$ Supported by Russian Foundation for Basic Research
      grant no. 96-02-00019 \\
% ...
$ ^l$ Partially Supported by Russian Foundation
      for Basic Research, grant    no. 00-15-96584 \\
}

\clearpage
\newpage

\section{Introduction}

One of the most remarkable results arising from the study of 
deep-inelastic $ep$ scattering (DIS) at HERA is the large range of 
squared four-momentum transfer $Q^2$ over which perturbative QCD 
calculations are able to describe the measurements of the inclusive 
cross section. Next-to-leading order (NLO) calculations are successful 
from values of $Q^2$ as low as a few GeV$^2$ up to $Q^2 \sim
35\,000\,$GeV$^2$~\cite{Adloff:2001qk,Breitweg:2000yn,Adloff:2001qj,
Breitweg:1999id,Adloff:2000ah}. Investigations of the hadronic
final state have shown that QCD is also able to describe events
containing two highly energetic
jets~\cite{Adloff:2001hm,Adloff:2001tq,Breitweg:2001rq}. Such
investigations have tended to require large inter-jet separations,
that is, a large relative jet transverse momentum, or a large
transverse jet energy in the Breit frame~\cite{Breit,Adloff:2001tq}. 
Typically only about a tenth of the DIS sample is then classified as dijet
events. These large scales are chosen to avoid the region in which
multiple parton emission is likely to become significant. Here, we
examine the possibility that fixed order perturbative QCD is able
to describe the hadronic final state in DIS even where jet
separations are small. Some hints that this may be possible have
been seen in measurements of event shape
variables~\cite{Adloff:2000gn,Breitweg:1998ug}.

The study proceeds by first identifying the minimum inter-jet
separation for which the rate of dijet production is successfully
described by NLO QCD calculations. Using this separation, about
$1/3$ of DIS events are classified as dijet. Based on this sample 
a number of measurements is made and compared with
perturbative QCD. At the lower end of the $Q^2$ range studied,
\mbox{$Q^2 \ge 150\,$GeV$^2$}, the sample is dominated by gluon induced
events, $eg \rightarrow eq\overline{q}$, whereas at the upper end
quark induced events predominate, $eq \rightarrow eqg$. The data
thus provide a thorough test of the QCD calculations.

An alternative QCD based description of the measurements is
provided by Monte Carlo models which describe DIS using leading
order (LO) QCD matrix elements matched to parton showers. The
inclusion of the latter would suggest that these might describe
dijet data in the region in which jet separations are very small.
Indeed, studies have been made of sub-jet multiplicities and jet
shapes, in which Monte Carlo models incorporating parton showers
have performed reasonably
well~\cite{Adloff:1999ni,Breitweg:1999js}. The LO calculations
incorporated in these models should ensure they are also able to
describe data at large jet separations. However, their overall
performance in describing the hadronic final state in DIS has not
yet been satisfactory \cite{Brook:1998jd}. In the present
analysis, we confront our measurements with the QCD model
RAPGAP~\cite{Jung:1995gf} which was not considered in
\cite{Brook:1998jd}.

\section{Experimental procedure}
\subsection{Selection of DIS events}

The present analysis is based on a data sample corresponding to an
integrated luminosity of \mbox{$\sim$ 35~pb$^{-1}$} recorded in 1995-97
with the H1 detector at HERA. In this period HERA operated with
positron and proton beams of 27.5 GeV and 820 GeV energy, respectively, 
yielding a centre-of-mass energy $\sqrt{s}$ of 300 GeV.

A detailed description of the H1 detector is given in
\cite{Abt:1997hi}. The detector components of most importance for
this study are the central tracking system and the liquid argon
calorimeter. We use a coordinate system with its origin at the
nominal interaction point and its positive $z$ axis along the
direction of the outgoing proton beam. Polar angles are denoted by
$\theta$ and the ``forward'' region is that with $\theta<90^{\circ}$.
Neutral current DIS events are selected using criteria similar to
those described in \cite{Adloff:2000ah}. These include the
requirement that a scattered positron be identified in the liquid
argon calorimeter at a polar angle $\theta_{e} < 150^\circ$. The
positron reconstruction method and the fiducial cuts on the
positron impact position in the liquid argon calorimeter are
described in \cite{Adloff:2000ah}. The value of $Q^{2}$,
determined from the energy and polar angle of the scattered
positron, must exceed $150\,$GeV$^{2}$. The Bjorken scaling variable 
$y$ must satisfy $0.1 < y < 0.7$. At low $y$, measurements of $y$ using 
the polar angles of the positron and of the reconstructed hadronic 
final state \cite{Adloff:2001qk}, $y_{da}$, are more precise than those 
using the energy and polar angle of the scattered positron, $y_e$. At high 
$y$ the situation is reversed. The low $y$ requirement is thus applied using 
the double angle measurement, $0.1 < y_{da}$, whereas the high $y$ 
restriction is applied using the positron measurement, $y_e < 0.7$. 
The selection yields a sample of $\sim 60\,000$ DIS events with negligible
background \cite{Adloff:2000ah}.

\subsection{Jet algorithm and observables}

Jets are reconstructed with the modified Durham algorithm,
described in more detail in~\cite{Catani:1991hj,Adloff:2001hm},
which is applied in the laboratory frame. Hadronic energy deposits
measured in the liquid argon calorimeter and the backward
``SpaCal'' calorimeter are used, as are tracks reconstructed in
the central tracking chambers, avoiding double counting of energy. 
All of these are referred to as ``proto-jets'' in the following and 
are required to have a polar angle greater than 7$^{\circ}$ to ensure 
that they are well measured. The proton remnant, which escapes direct 
detection, is included in the jet reconstruction by forming a 
missing-momentum four-vector, which is treated as an additional proto-jet.

The algorithm uses the relative $k_{T\, ij}^{2}= 2 \min[E_{i}^2,
E_{j}^2]\,(1-\cos\theta_{ij})$ of proto-jets $i$,~$j$ as a measure
of their separation, where $E_{i}$ and $E_{j}$ are the energies of
the proto-jets $i$ and $j$, and $\theta_{ij}$ the angle between
them. The pair $i$, $j$ with the minimum $k_{T\,ij}$ is combined
to form a new proto-jet by adding the four-momenta $p_{i}$ and
$p_{j}$. The iterative clustering procedure is ended when exactly
two final state jets and the proton remnant jet remain.

In order to select a sample of dijet events we define the variable $y_2$:
\begin{equation} y_2= \frac{\dps \min_{i,j,i\not=j} k_{T\,ij}^2}
                                 {\dps W^2},
\end{equation}
where $i,j$ may be any of the two final state jets or the remnant
jet, and $W$ is the invariant mass of all objects entering the jet
algorithm, including the missing-momentum vector. It is ensured
that the jets are well contained within the liquid argon
calorimeter by requiring that, for both non-remnant jets,
$10^\circ < \theta_{\textrm{jet}} < 140^\circ $.

We study the following observables: $y_2$ as defined above; the
polar angles of the forward and the backward (non-remnant) jets in
the laboratory frame ${\theta_{\textrm{fwd}}}$ and ${\theta_{\textrm{bwd}}}$; the
dimensionless variables ${x_p}$ and ${z_p}$; and the average transverse 
energy of the two final state (non-remnant) jets in the Breit frame 
${\overline{E}_{T\,{\rm Breit}}}$. In order to calculate 
$\overline{E}_{T\,{\rm Breit}}$, the jets found in the laboratory frame 
are boosted into the Breit frame\footnote{The Breit frame is related to the
hadronic centre-of-mass frame by a longitudinal boost. In both
frames, the total transverse momentum of the hadronic final state
is zero whereas in the laboratory frame it is constrained to
balance the scattered lepton's transverse momentum.}. We calculate
$z_p$ and $x_p$ according to $$z_{p} \equiv
 \displaystyle\frac{\min_{i = 1,2}  E_{i}\,(1 - \cos\theta_{i})}
{\sum_{i = 1,2} \,\displaystyle E_{i}\,(1 -
\cos\theta_{i})}\,\,\,\, {\rm and} \,\,\,\, x_{p} \equiv
\frac{\displaystyle Q^2}{\displaystyle Q^2 + m_{12}^2} \, ,$$
where $E_{i}$ and $\theta_{i}$ are the energies and polar angles
of the two (non-remnant) jets, and $m_{12}$ is the invariant dijet mass. 
Matrix elements for dijet production are frequently expressed
using these variables \cite{Korner:1989bp,Seymour:1995we,Mirkes:1997uv}. 
In leading order QCD, the cross section diverges for $z_p\rightarrow 0$ and
$x_p\rightarrow 1$ due to collinear and infrared singularities.


In the QCD models and the NLO calculations, jets are defined by
applying the above algorithm to the  four-momenta of hadrons or
partons. In particular, the requirement that the polar angle be
greater than $7^{\circ}$ is always applied.

\subsection{Data correction and systematic uncertainties}

The procedures used for data correction and the determination of
the systematic uncertainties are similar to those described
in~\cite{Adloff:2001hm,Adloff:2001tq}. The measured jet
distributions are corrected for the effects of the limited
detector acceptance and resolution, and for the effects of QED
radiation. This is done using bin-by-bin correction factors
determined with the QCD Monte Carlo models
ARIADNE~\cite{Lonnblad:1992tz} and LEPTO~\cite{Ingelman:1997mq},
both of which are incorporated in DJANGO~\cite{Schuler:1991yg}.
The average of the measured jet distributions corrected with
ARIADNE or LEPTO is taken as the final result. As a cross check,
some distributions are also corrected using a regularized
unfolding technique~\cite{Blobel:1984ku} (for a brief description
see~\cite{Adloff:1998ss}). The two correction methods lead to very
similar results.

The dominant systematic errors are due to the model dependence of
the corrections and the uncertainty of the electromagnetic and
hadronic energy scales of the calorimeters. The errors are added 
in quadrature to yield the total systematic error. The model uncertainty 
is taken to be the difference between the averaged correction factors 
and those determined with a single model, and is of the order of $5\%$. 
The uncertainty of the electromagnetic energy scale of the liquid argon
calorimeter ranges from $\pm 0.7\%$ to $\pm 3\%$, depending on the scattered 
positron's impact position. The changes in the measured dijet distributions 
resulting from the variation of this scale within its uncertainty lead to 
a systematic error of less than $1\%$. The hadronic energy scale of the 
liquid argon calorimeter is varied by $\pm 4\%$, which leads to an average
uncertainty of $4\%$ in the dijet measurements.

\subsection{Perturbative QCD and model calculations}

The perturbative QCD predictions presented in the following are
calculated using the DISENT program~\cite{Catani:1997vz}. The
agreement of DISENT with other NLO programs is discussed
in~\cite{Graudenz:1998jg,Duprel:1999wz,McCance:1999jh,Antonelli:2000kx}.
In the calculations, we use the CTEQ5M parton density
functions~\cite{Lai:2000wy}, choose $Q$ as the renormalization and
factorization scale if not otherwise stated and set the value of
$\alpha_s(M_Z)$ to 0.1183. This gives a good description of the
inclusive DIS cross section in the kinematic range of this
analysis~\cite{Adloff:2000ah}. Other choices of recent parton density
parameterizations, for example those determined by the H1
collaboration in~\cite{Adloff:2000ah}, are found to yield very
similar NLO predictions. The size of the hadronization effects is
determined using the QCD Monte Carlo models LEPTO and ARIADNE.
Hadronization correction factors are obtained by dividing the jet
distributions for hadrons by those determined from the partons
after the parton shower or dipole cascade, respectively. The
average of the hadronization corrections obtained from the two
models is applied to the NLO calculations in the comparisons. The
uncertainty of this procedure is conservatively estimated to be
half the size of the correction. This is significantly larger than
the difference between the corrections determined with ARIADNE and
LEPTO.

Comparisons are also made with the QCD based Monte Carlo program 
RAPGAP. This models QCD radiation with initial and final state
parton showers~\cite{Bengtsson:1988rw} combined with leading order
QCD matrix elements~\cite{Korner:1989bp,Seymour:1995we,Mirkes:1997uv}.
Hadronization is simulated using the Lund string
model~\cite{Andersson:1983ia,Sjostrand:1987hx}. We use the default
model parameters\footnote{The cut-off parameter for the LO matrix 
element calculation PT2CU is set to 5 GeV$^2$.} and the CTEQ4L parton
density functions~\cite{Lai:1997mg}.

\section{Determination of minimum required jet separation}

\begin{figure}[h!]\vspace*{-1cm}
\begin{center}
\epsfig{file=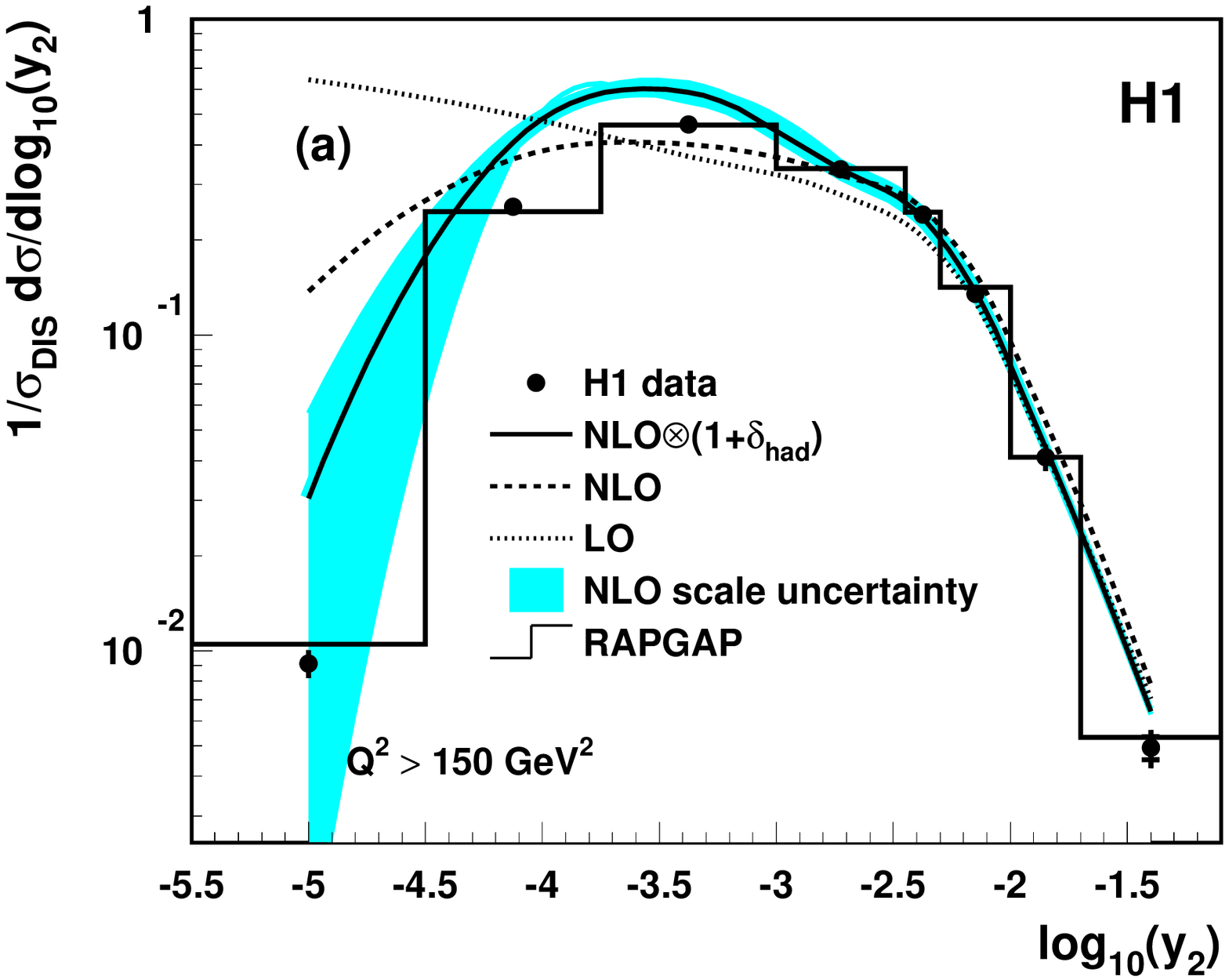,width=14.5cm,clip} 
\epsfig{file=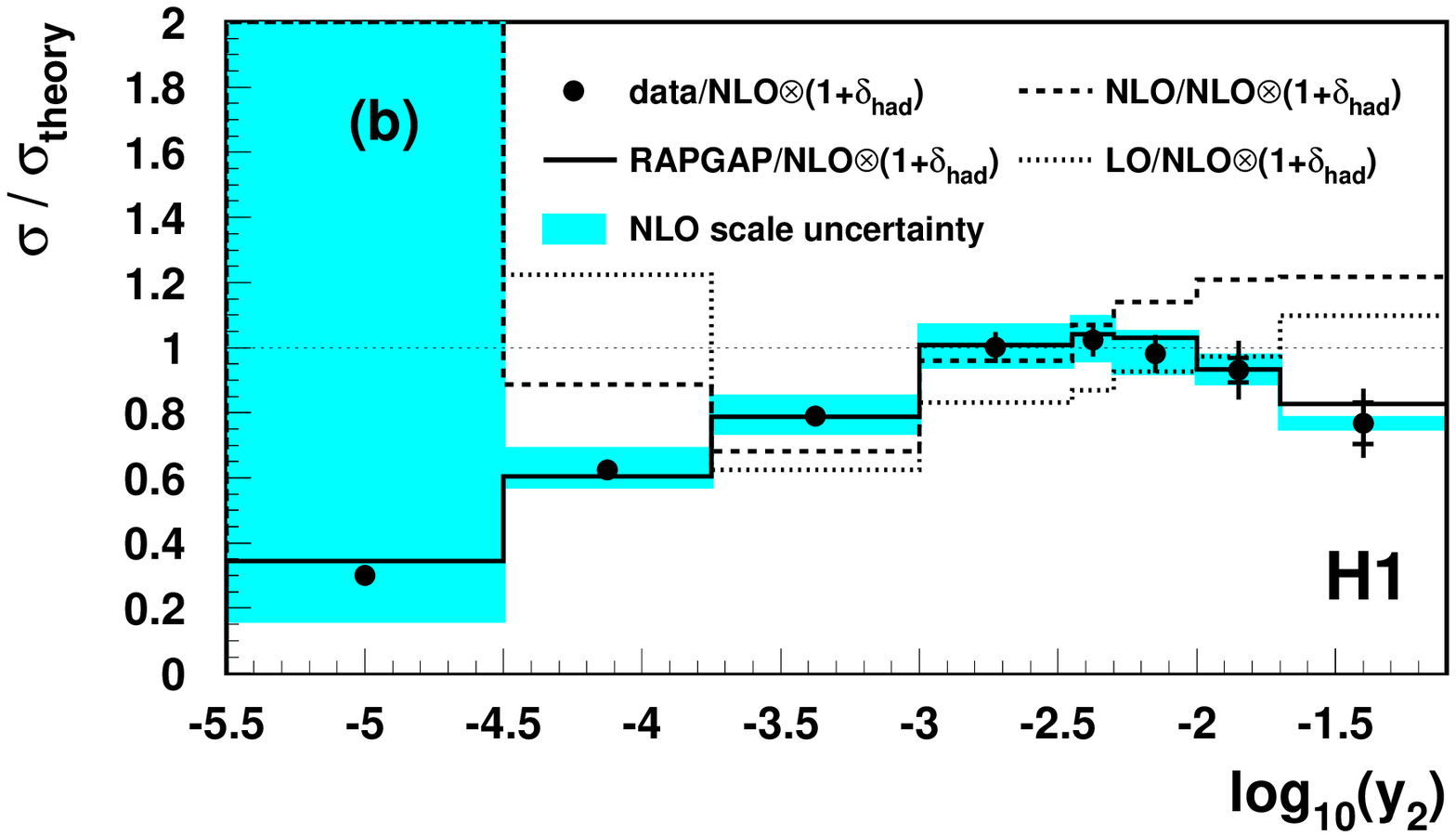,width=14.5cm,clip} 
\caption{(a) Distribution of $y_2$ for $Q^2 > 150\,$GeV$^2$, 
determined with the modified Durham algorithm. The events satisfy
$10^{\circ}<\theta_{\rm jet}<140^{\circ}$. Here, and in the following
figures, the statistical errors are given by the inner error bars
and the outer error bars correspond to the quadratic sum of the
statistical and systematic errors.  Also shown are perturbative
QCD calculations in LO, in NLO with and without hadronization
corrections, and the predictions of the QCD model RAPGAP. The
shaded band shows the renormalization scale uncertainty of the NLO
calculations, which is estimated here and below by varying $\mu_R$
in the range $Q/2 $ to $2\,Q$.
\newline
(b) The ratios of the data and various predictions. The vertical error
bars correspond to the uncertainty of the data only.}
\label{dh.logy2}
\end{center}
\end{figure}

A direct way of determining the minimum jet separation
necessary to ensure that NLO QCD describes the dijet production
rate is to compare measurements of the jet separation itself with
calculations. The measured $y_2$ distribution, normalized to the
inclusive DIS cross section $\sigma_{DIS}$ in the region defined 
by $Q^2 > 150\,$GeV$^2$, $0.1 < y < 0.7$ and $\theta_e < 150^\circ$, 
is shown with the results of various calculations 
in Figure~\ref{dh.logy2} and listed in Table~\ref{tab.y2}.

We observe that the NLO perturbative QCD calculations combined
with hadronization corrections overestimate the measured cross
section drastically in the region $y_2 < 0.001$, where jet
separations are smallest. Here the difference between LO and NLO
predictions is large. The renormalization scale dependence,
estimated by varying $\mu_R$ in the range $Q/2 < \mu_R <2\, Q$,
and the hadronization corrections are also large. All three
criteria suggest that fixed order perturbative QCD predictions are
not reliable in the region $y_2 < 0.001$, and agreement of the
calculations with the data cannot be expected. The situation is
much improved at $y_2 > 0.001$, and a good description of the data
is observed from $y_2 \sim 0.001$ up to the largest $y_2$ values,
where jet structures are most distinct. (The deviation of NLO QCD in 
the highest $y_2$ bin can be explained by the exceptionally large
parton density function dependence in this region of the dijet
phase space.)

The $y_2$ distribution of Figure~\ref{dh.logy2} is also compared
to the QCD model RAPGAP, which describes the $y_2$ cross section
over the full measured range. In particular, the region of very
low $y_2$ is well described, which suggests that the combination of parton 
showers and Lund string hadronization used in RAPGAP accurately models 
multi-parton emissions.

\section{Study of dijet sample}

Motivated by the agreement of NLO perturbative QCD with the data
at $y_2 > 0.001$ in Figure~\ref{dh.logy2}, we investigate this
sample of dijet events
in more detail.
This sample contains about $1/3$ of the selected DIS events.

\begin{figure}[h]\vspace*{5.cm}
\begin{picture}(180,150)
\epsfig{file=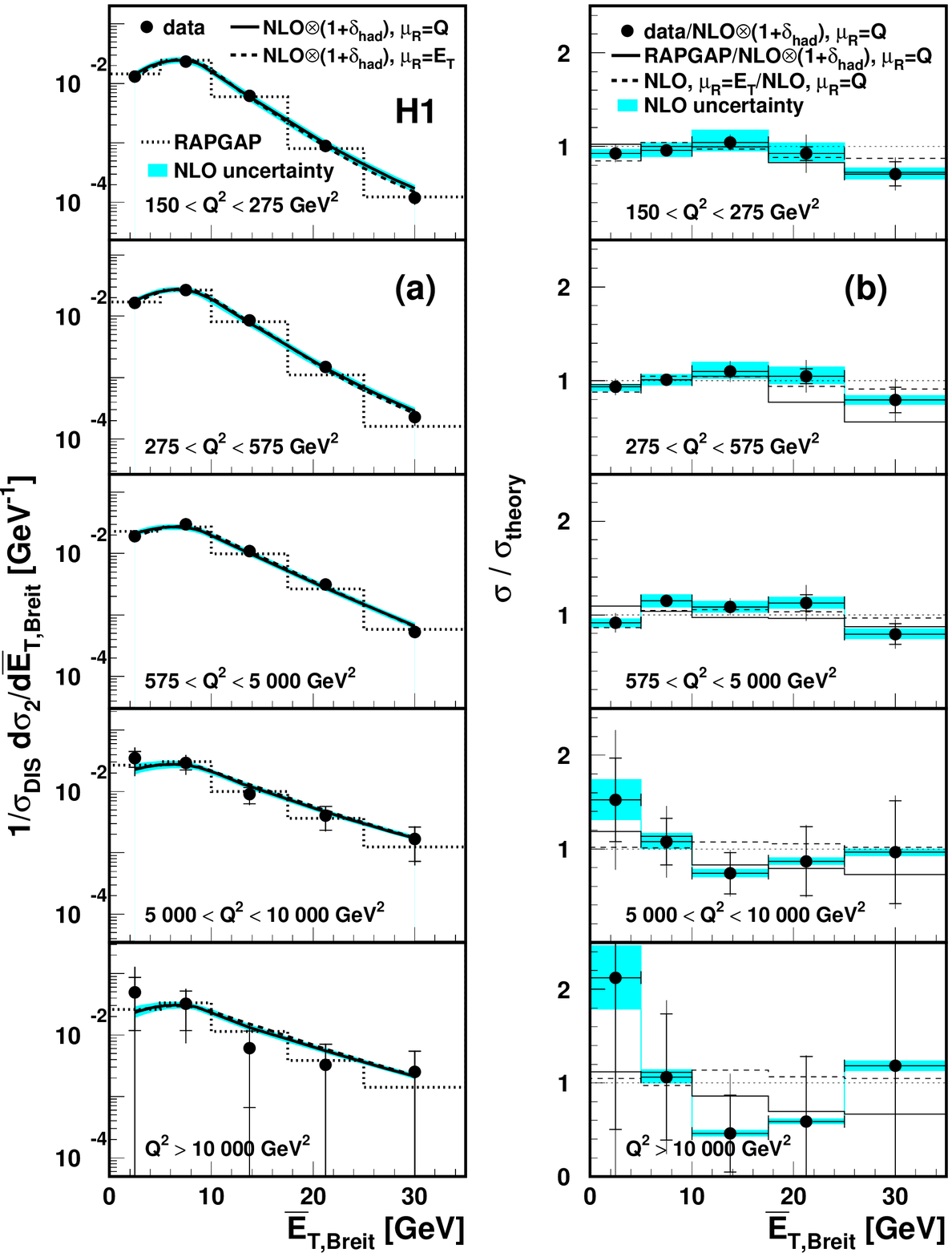, height=21.cm,clip}
\end{picture}
\vspace{-0.7cm} \caption{(a) Dijet event $\overline{E}_{T\,{\rm
Breit}}$ distributions determined with the modified Durham
algorithm in various $Q^2$ ranges. The dijet events satisfy $y_2
> 0.001$ and $10^{\circ}<\theta_{\rm jet}<140^{\circ}$. Also shown are
perturbative QCD calculations in NLO with $\mu_R=Q$ and
$\mu_R=\overline{E}_{T\,{\rm Breit}}$, and the predictions of the
QCD model RAPGAP. The shaded band corresponds to the quadratic sum of the 
hadronization and renormalization scale uncertainties.
\newline
(b) The ratios of the data and various predictions. The vertical 
error bars correspond to the uncertainty of the data only.} 
\label{dh.001.et}
\end{figure}

The dijet cross section $d\sigma_2/d{\overline{E}_{T\,{\rm Breit}}}$ in 
several ranges of $Q^2$ is shown in Figure~\ref{dh.001.et} and listed in
Table~\ref{tab.et}. The normalization is again to $\sigma_{DIS}$, the DIS cross
section in the region defined by $0.1 < y < 0.7$, $\theta_e <
150^\circ$ and the indicated $Q^2$ range. For a sizable 
fraction of the events, $\overline{E}_{T\,{\rm Breit}}$ is smaller than
$5\,$GeV and the mean value of $\overline{E}_{T\,{\rm Breit}}$
over the dijet sample is $\sim 6\,$GeV. The more restrictive dijet
samples used in other QCD analyses typically require that one of the jets
have $E_{T\,{\rm {\rm Breit}}} > 7.5\,$GeV or
higher~\cite{Adloff:2001tq,Breitweg:2001rq}. We compare the
measurements with perturbative QCD calculations in NLO for two
choices of renormalization scale, $\mu_R=Q$ and $\mu_R
={\overline{E}_{T\,{\rm Breit}}}$.\footnote{For the choice
$\mu_R={\overline{E}_{T\,{\rm Breit}}}$, the cut
$\overline{E}_{T\,{\rm Breit}}>0.5$ GeV is applied to improve the
convergence of the DISENT calculations. This has a negligible effect
in the selected dijet phase space.} Perturbative QCD in
NLO describes the $\overline{E}_{T\,{\rm Breit}}$ distributions
well, including the region $\overline{E}_{T\,{\rm Breit}} <
5\,$GeV. Although the numerical values of
$Q^2$ and $\overline{E}^2_{T\,{\rm Breit}}$ are very different for
most events, the difference between the NLO predictions for
the different scales is small. RAPGAP also describes the ${\overline{E}_{T\,{\rm
Breit}}}$ distributions well.

\begin{figure}[h]\vspace*{5.cm}
\begin{picture}(180,150)
\epsfig{file=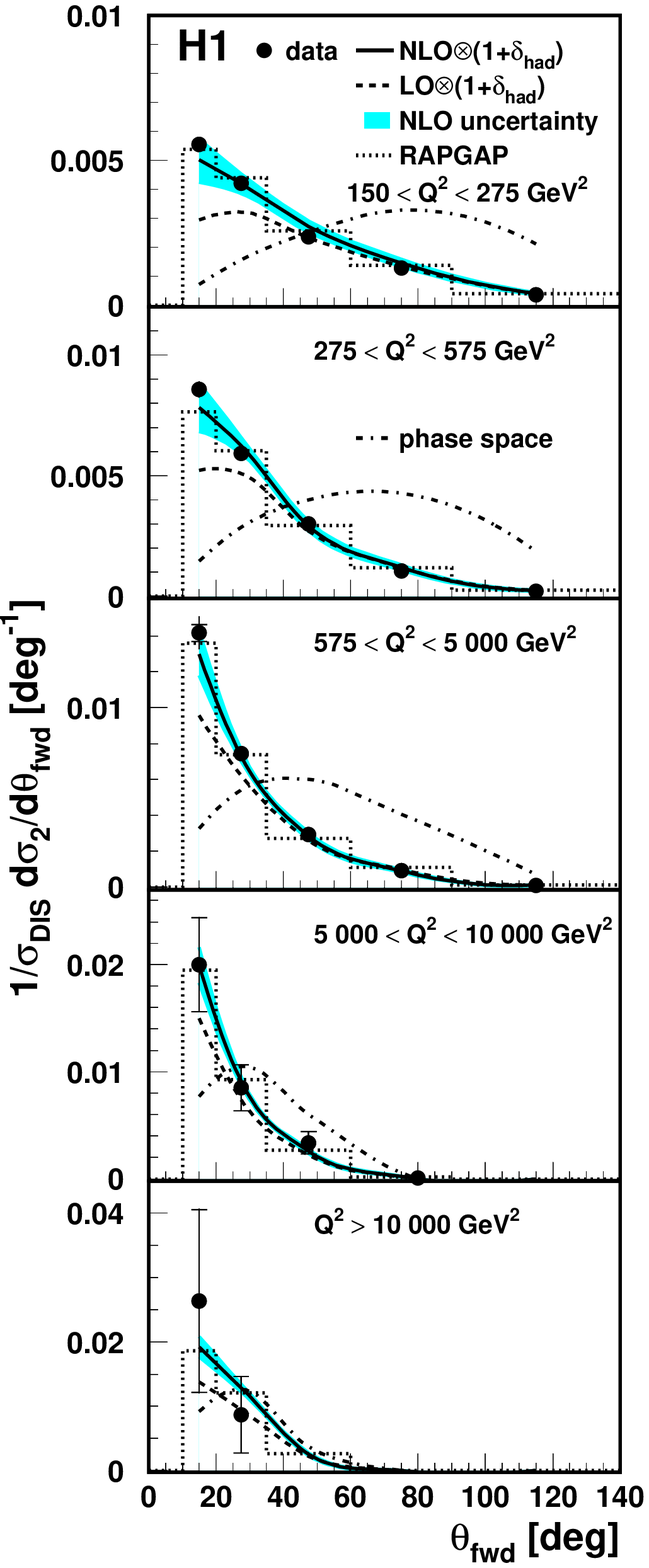, height=21.cm,clip}
\epsfig{file=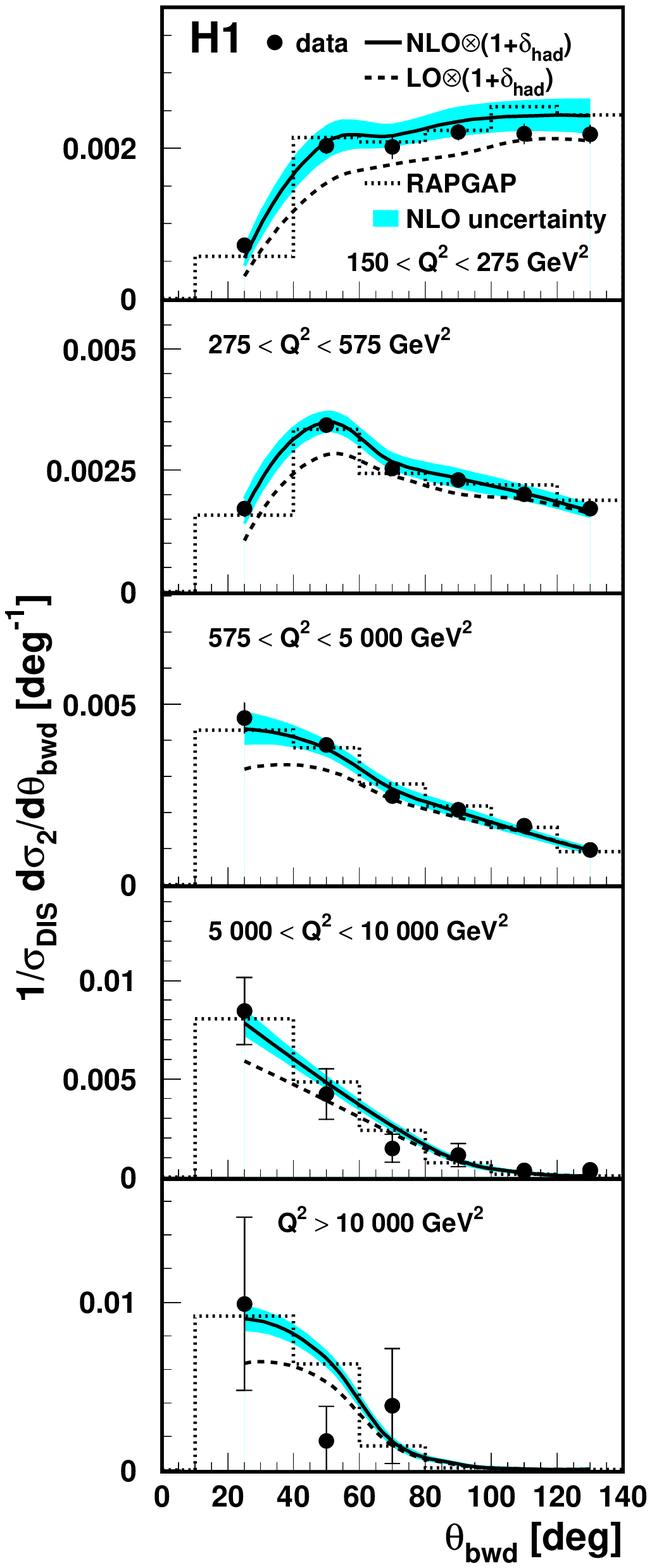, height=21.cm,clip}
\end{picture}
\caption{Dijet event $\theta_{\textrm{fwd}}$ and $\theta_{\textrm{bwd}}$
distributions determined with the modified Durham algorithm in
various $Q^2$ ranges. The dijet events satisfy $y_2 > 0.001$ and
$10^{\circ}<\theta_{\rm jet}<140^{\circ}$. Also shown are perturbative
QCD calculations in NLO and LO, the predictions of the QCD
model RAPGAP, and a phase space calculation in arbitrary normalization. The 
shaded band corresponds to the quadratic sum of the hadronization and renormalization 
scale uncertainties.} 
\label{dh.001.tfwtbw}
\end{figure}

The measured distributions of the forward and backward
jet polar angles, $\theta_{\textrm{fwd}}$ and $\theta_{\textrm{bwd}}$, are shown in
Figure~\ref{dh.001.tfwtbw} (and Table \ref{tab.th}) in five $Q^2$ ranges. 
The $\theta_{\textrm{fwd}}$ distributions increase strongly towards small
angles and are less dependent on $Q^2$ than the $\theta_{\textrm{bwd}}$ 
distributions, as expected if the forward jets are largely due to initial 
state radiation off the constituents of the proton. The distributions are well 
described by NLO perturbative QCD and by the RAPGAP model. In particular 
at small jet polar angles and at relatively small $Q^2$, the LO calculations 
differ considerably from the NLO ones and are unable to describe the data.
A prediction including only the phase space contribution, without the QCD
matrix elements, results in relatively flat $\theta_{\textrm{fwd}}$
distributions and does not describe the data.

The $\theta_{\textrm{bwd}}$ distribution has its maximum at large polar
angles in the kinematic region $150 < Q^2 < 275$ GeV$^2$. With
increasing $Q^2$, the maximum shifts into the forward region of
the detector. Again, this is as expected due to the increasing
fraction of the proton's momentum transferred to the 
hadronic final state as $Q^2$ grows. Perturbative QCD in NLO and the QCD 
model RAPGAP describe the distributions well, while the LO QCD predictions
fail.

\begin{figure}[h]\vspace*{5.cm}
\begin{picture}(180,150)
\epsfig{file=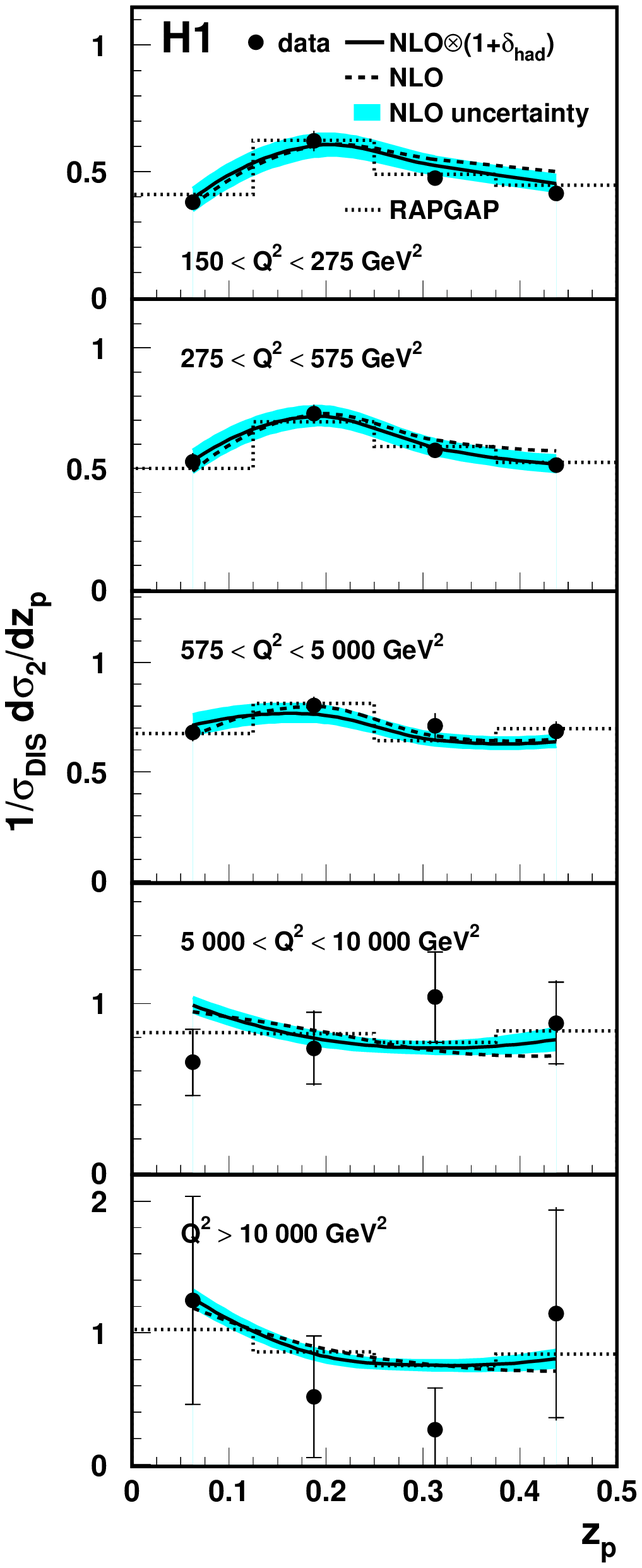, height=21.cm,clip}
\epsfig{file=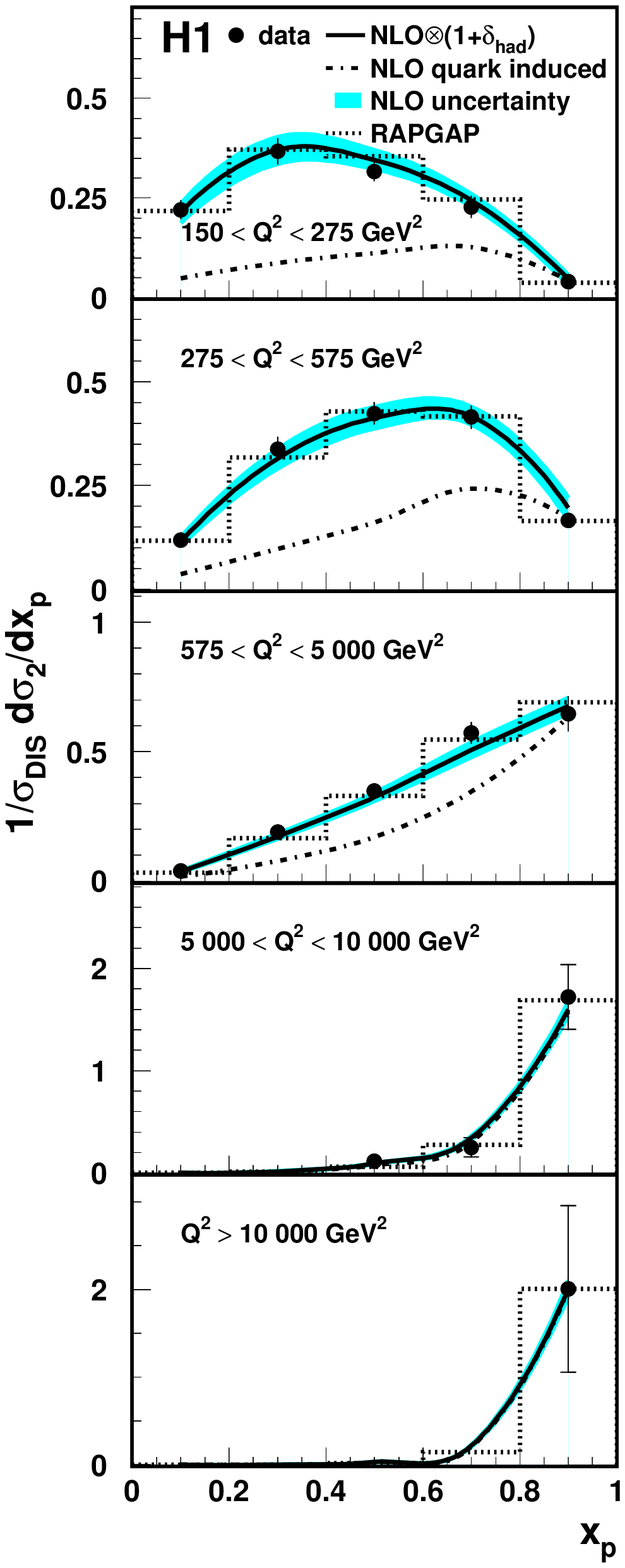, height=21.cm,clip}
\end{picture}
\caption{Dijet event $z_p$ and $x_p$ distributions determined with
the modified Durham algorithm in various $Q^2$ ranges. The dijet
events satisfy $y_2 > 0.001$ and
$10^{\circ}<\theta_{\rm jet}<140^{\circ}$. Also shown are perturbative
QCD calculations in NLO with and without hadronization
corrections, and the predictions of the QCD model RAPGAP. The shaded band 
corresponds to the quadratic sum of the hadronization and renormalization scale 
uncertainties. Note that the quark-induced contribution to the dijet cross 
section is close to 100\% for $Q^2 > 5000$ GeV$^2$.} 
\label{dh.001.zpxp}
\end{figure}

The $x_p$ and $z_p$ distributions are shown in Figure~\ref{dh.001.zpxp} 
and listed in Table \ref{tab.zpxp}. The $z_p$ distribution is well described 
by both NLO QCD and RAPGAP. The NLO calculations without hadronization 
corrections are also shown. They describe the data fairly well since the 
hadronization corrections are small. (Note that, by definition, the error 
band includes half of the hadronization corrections.) The $x_p$ distribution 
is also well described by NLO QCD. The NLO predictions including only the 
quark contribution ($eq \rightarrow eqg$) to the dijet cross section are
shown separately. The proportion of the dijet cross section  that
is quark-induced is expected to be largest at large $x_p$ values,
and varies from $\approx$ 30\% at the lowest $Q^2$ range to nearly
100\% at $Q^2 > 10\,000$ GeV$^2$. This illustrates that our
measurements are sensitive tests of both the gluon- and the quark-initiated 
NLO matrix elements.

\section{Summary}

We have investigated the minimum inter-jet separation necessary to
ensure that next-to-leading order QCD calculations are able to
accurately describe dijet production in deep-inelastic scattering.
Using data in the kinematic range $150 < Q^2 < 35\,000$ GeV$^2$,
the required separation is found to be small, resulting in the
selection of a dijet sample containing about $1/3$ of the DIS
events, a significantly larger proportion than the approximately $1/10$
obtained with more typical jet selection criteria. Measurements
of the distribution of variables sensitive to the dynamics of jet
production are well described by NLO QCD calculations, for either
choice of renormalization scale $Q$ or $\overline{E}_{T\,{\rm Breit}}$.  
This good description extends to unexpectedly small jet separations 
and covers regions in which both gluon and quark induced processes 
dominate.

Due to their precision and to the large phase space covered, the
measurements also significantly constrain QCD Monte Carlo models.
A good description of the data may be achieved by models which
combine leading order QCD matrix elements with parton showers, as
demonstrated here in the case of RAPGAP.

\section*{Acknowledgements}

We are very grateful to the HERA machine group whose outstanding
efforts made this experiment possible. We acknowledge the support
of the DESY technical staff. We appreciate the big effort of the
engineers and technicians who constructed and maintain the
detector. We thank the funding agencies for financial support of
this experiment. We wish to thank the DESY directorate for the
support and hospitality extended to the non-DESY members of the
collaboration.

\clearpage\newpage

 \begin{table}[h!]
 \begin{center}
 \begin{tabular}{|c||c|r|r| }
 \hline
 $-\log_{10} (y_2)$
&$\frac{\dps 1}{\dps \sigma_{DIS}}\frac{\dps {\rm d}\sigma}{\dps
{\rm d}\log_{10} y_2}$ &$\delta_{stat}$(\%)&$\delta_{sys}$(\%)
\\ \hline
  & \multicolumn{  3}{|c|}{$150<Q^2<5000$ GeV$^2$} \\
 \hline
$5.5 - 4.5$& 0.00912 &$\pm$5.2&$\pm 8.8$
 \\
 \hline
$ 4.5 - 3.75$&  0.255 &$\pm$1.5&$\pm 4.1$
 \\ \hline
$3.75- 3.0$ & 0.465 &$\pm$1.1&$\pm 0.9$
 \\
 \hline
$3.0  - 2.45$ &   0.335 &$\pm$1.3&$\pm 4.6$
 \\
 \hline
$2.45 - 2.3$ &  0.241 &$\pm$2.5&$\pm 4.3$
 \\
 \hline
$2.3  - 2.0$& 0.135 &$\pm$2.4&$\pm 5.4$
 \\
 \hline
$2.0  - 1.7$& 0.0411 &$\pm$4.1&$\pm 8.7$
 \\
 \hline
$1.7  - 1.1$& 0.0049 &$\pm$8&$\pm 11$
 \\
 \hline
 \end{tabular}
 \end{center}
\caption{Normalized jet cross sections as a function of $y_{2}$
for $Q^2 > 150$ GeV$^2$, $\theta_{e}$ $<150^{\circ}$,  $0.1< y<
0.7$, and $10^{\circ} < \theta_{\rm jet} < 140^{\circ}$. The relative
statistical errors $\delta_{stat}$ and systematic errors
$\delta_{sys}$ are given in per cent.} \label{tab.y2}
 \end{table}


 \begin{table}[h!]
 \begin{center}
 \begin{tabular}{|c||c|r|r||c|r|r|  }
 \hline
%
%
 $\overline{E}_{T\,{\rm Breit}}$
&$\frac{\dps 1}{\dps \sigma_{DIS}}\frac{\dps {\rm
d}\sigma_{2}}{\dps {\rm d}\overline{E}_{T\,{\rm Breit}}}$
&$\delta_{stat}$(\%)&$\delta_{sys}$(\%) &$\frac{\dps 1}{\dps
\sigma_{DIS}}\frac{\dps {\rm d}\sigma_{2}}{\dps {\rm
d}\overline{E}_{T\,{\rm Breit}}}$
&$\delta_{stat}$(\%)&$\delta_{sys}$(\%) \\
{\small [GeV]}& {\small[GeV$^{-1}$]} & & & {\small [GeV$^{-1}$]} & & \\
\hline
  & \multicolumn{  3}{|c||}{$150<Q^2<275$ GeV$^2$} & \multicolumn{ 3}{|c|}{$275<Q^2<575$ GeV$^2$}\\
 \hline
$0 - 5$&  0.013 &$\pm$2.5&$\pm 9.9$ &  0.0165 &$\pm$2.9&$\pm 8.8$
 \\
 \hline
$5- 10$&  0.0233 &$\pm$2.2&$\pm 5.9$ & 0.027 &$\pm$3&$\pm 21$
 \\
 \hline
$10 - 17.5$ & 0.00619 &$\pm$3.3&$\pm 7.4$ &  0.00851
&$\pm$3.6&$\pm 9.7$
 \\
 \hline
$17.5 - 25$ & 0.00089 &$\pm$8 &$\pm 21$ & 0.0015 &$\pm$8&$\pm 15$
 \\
 \hline
$25 - 35$ & 0.00012 &$\pm$18 &$\pm 17$ &   0.00023 &$\pm$17&$\pm
23$
 \\
 \hline
 \hline
  & \multicolumn{  3}{|c||}{$575<Q^2<5000$ GeV$^2$} & \multicolumn{ 3}{|c|}{$5000<Q^2<10\,000$ GeV$^2$}\\
 \hline
$0 -5$& 0.0193 &$\pm$4&$\pm 10$ &  0.035 &$\pm 29$&$\pm 39$
 \\
 \hline
$5 - 10$& 0.0301 &$\pm$3.3& $\pm 4.6$ & 0.029 &$\pm 23$&$\pm$27
 \\
 \hline
$10 - 17.5$ &   0.0110 &$\pm$4.4&$\pm 7.3$ & 0.0089 &$\pm$30&$\pm
15$
 \\
 \hline
$17.5 - 25$ &  0.0031 &$\pm$8&$\pm 15$ &  0.0040 &$\pm$42&$\pm 14$
 \\
 \hline
$25 - 35$ & 0.00053 &$\pm$14&$\pm 14$ &  0.0017 &$\pm$57&$\pm$ 26
 \\
 \hline
 \hline
  & \multicolumn{  3}{|c||}{$Q^2> 10\,000$ GeV$^2$} & \multicolumn{ 3}{c}{}\\\cline{1-4}
 $0 - 5$&   0.049
&$\pm 76$&$\pm 100$ &\multicolumn{ 3}{c}{}\\\cline{1-4} $5 - 10$&
0.032 &$\pm 63$& $\pm 44$ &\multicolumn{ 3}{c}{}\\\cline{1-4} $10
- 17.5$ &  0.0061 &$\pm 89$&$\pm 100$ &\multicolumn{
3}{c}{}\\\cline{1-4} $17.5 - 25$ &  0.0032 &$\pm100$&$\pm 23$
&\multicolumn{ 3}{c}{}\\\cline{1-4} $25 - 35$ &  0.0025 &$\pm
100$&$\pm 19$ &\multicolumn{ 3}{c}{}\\\cline{1-4}
 \end{tabular}
 \end{center}
\caption{Normalized dijet event cross sections as a function of
$\overline{E}_{T\,{\rm Breit}}$, determined with the modified
Durham algorithm. The selection criteria are $\theta_{e}$
$<150^{\circ}$, $0.1< y< 0.7$, $y_2 > 0.001$ and $10^{\circ} <
\theta_{\rm jet} < 140^{\circ}$. The $Q^2$ range is given in the
table. The relative statistical errors $\delta_{stat}$ and
systematic errors $\delta_{sys}$ are given in per cent.}
\label{tab.et}
 \end{table}

\clearpage\newpage
 \begin{table}[h!]
 \begin{center}
 \begin{tabular}{|c||c|r|r||c|r|r|  }
 \hline
 $\theta_{\textrm{fwd}}$
&$\frac{\dps 1}{\dps \sigma_{DIS}}\frac{\dps {\rm
d}\sigma_{2}}{\dps {\rm d}\theta_{\textrm{fwd}}}$
&$\delta_{stat}$(\%)&$\delta_{sys}$(\%) &$\frac{\dps 1}{\dps
\sigma_{DIS}}\frac{\dps {\rm d}\sigma_{2}}{\dps {\rm
d}\theta_{\textrm{fwd}}}$ &$\delta_{stat}$(\%)&$\delta_{sys}$(\%)
\\
\small{ [deg]}& \small{[deg$^{-1}$]} & & & \small{[deg$^{-1}$]} & & \\
\hline
  & \multicolumn{  3}{|c||}{$150<Q^2<275$ GeV$^2$} & \multicolumn{ 3}{|c|}{$275<Q^2<575$ GeV$^2$}\\
 \hline
$10 - 20$&  0.00555 &$\pm$2.8&$\pm 2.6$ & 0.00858 &$\pm$2.9&$\pm
2.5$
 \\
 \hline
$20 - 35$&    0.00422 &$\pm$2.6&$\pm 2.5$ &0.00592 &$\pm$2.9&$\pm
3.8$
 \\
 \hline
$35 - 60$ &  0.00236 &$\pm$2.9&$\pm 7.9$ & 0.00299 &$\pm$3.3&$\pm
3.3$
 \\
 \hline
$60 - 90$ &  0.00129 &$\pm$3.7&$\pm 5.8$ &  0.00105 &$\pm$4.9&$\pm
4.5$
 \\
 \hline
$90 - 140$ &  0.000364 &$\pm$5.1&$\pm 4.6$ &  0.00231
&$\pm$8.1&$\pm 6.2$
 \\
 \hline
 \hline
  & \multicolumn{  3}{|c||}{$575<Q^2<5000$ GeV$^2$} & \multicolumn{ 3}{|c|}{$5000<Q^2<10\,000$ GeV$^2$}\\
 \hline
$10 - 20$& 0.0142 &$\pm$3.4&$\pm 5.5$ &  0.020 &$\pm$22&$\pm 4$
 \\
 \hline
$20 - 35$& 0.00745 &$\pm$3.7& $\pm 3.9$ &  0.0085 &$\pm$25&$\pm 3$
 \\
 \hline
$35 - 60$ & 0.00292 &$\pm$4.7&$\pm 2.7$ &  0.0034 &$\pm$31&$\pm 6$
 \\
 \hline
$60 - 90$ &  0.000921 &$\pm$7.4&$\pm 4.8$ &  0.00017
&$\pm$100&$\pm 4$
 \\
 \hline
$90 - 140$ &  0.000146 &$\pm$15&$\pm 8.5$ & 0 & 
\multicolumn{2}{|c|}{$<0.0001~(68\%~\textrm{CL})$}
 \\
\hline
\hline
  & \multicolumn{  3}{|c||}{$Q^2> 10\,000$ GeV$^2$} & \multicolumn{ 3}{c}{}\\\cline{1-4}
 $10 - 20$&  0.026
&$\pm$54&$\pm 8$ &\multicolumn{ 3}{c}{}\\\cline{1-4} $20 - 35$&
0.0087 &$\pm$69& $\pm 3$ &\multicolumn{ 3}{c}{}\\\cline{1-4} $35 - 60$ & 
$0$ &\multicolumn{2}{|c||}{$<0.002~(68\%~\textrm{CL})$}&\multicolumn{ 3}{c}{}\\
\hline
\hline
\cline{1-7}
%
 $\theta_{\textrm{bwd}}$
&$\frac{\dps 1}{\dps \sigma_{DIS}}\frac{\dps {\rm
d}\sigma_{2}}{\dps {\rm d}\theta_{\textrm{bwd}}}$
&$\delta_{stat}$(\%)&$\delta_{sys}$(\%) &$\frac{\dps 1}{\dps
\sigma_{DIS}}\frac{\dps {\rm d}\sigma_{2}}{\dps {\rm
d}\theta_{\textrm{bwd}}}$ &$\delta_{stat}$(\%)&$\delta_{sys}$(\%)
\\
\small{ [deg]}& \small{[deg$^{-1}$]} & & & \small{[deg$^{-1}$]} & & \\
\hline
  & \multicolumn{  3}{|c||}{$150<Q^2<275$ GeV$^2$} & \multicolumn{ 3}{|c|}{$275<Q^2<575$ GeV$^2$}\\
 \hline
$10 - 40$&  0.000711 &$\pm$3.7&$\pm 3.9$ & 0.00171 &$\pm$3.2&$\pm
5.1$
 \\
 \hline
$40 - 60$&   0.00203 &$\pm$3.2&$\pm 4.6$ &0.00343 &$\pm$3.3&$\pm
3.2$
 \\
 \hline
$60 - 80$ &  0.00202 &$\pm$3.3&$\pm 7.2$ &0.00254 &$\pm$3.8&$\pm
3.7$
 \\
 \hline
$80 - 100$ &  0.00221 &$\pm$3.3&$\pm 3.7$ &  0.00230 &$\pm$4.0&$\pm
3.0$
 \\
 \hline
$100 - 120$ &  0.00219 &$\pm$3.3&$\pm 5.2$ &  0.00200
&$\pm$4.4&$\pm 4.3$
 \\
 \hline
$120 - 140$ & 0.00218 &$\pm$3.3&$\pm 4.6$ & 0.00171 &$\pm$4.6&$\pm
3.5$
 \\
 \hline
 \hline
  & \multicolumn{  3}{|c||}{$575<Q^2<5000$ GeV$^2$} & \multicolumn{ 3}{|c|}{$5000<Q^2<10\,000$ GeV$^2$}\\
 \hline
$10 - 40$& 0.00462 &$\pm$3.3&$\pm 8.6$ &  0.0085 &$\pm$20&$\pm 3$
 \\
 \hline
$40 - 60$& 0.00387 &$\pm$4.4& $\pm 2.6$ &  0.0042 &$\pm$30&$\pm 6$
 \\
 \hline
$60 - 80$ & 0.00245 &$\pm$5.4&$\pm 2.6$ &   0.0015 &$\pm$48&$\pm
16$
 \\
 \hline
$80 - 100$ & 0.00206 &$\pm$6.0&$\pm 3.6$ &  0.0011 &$\pm$52&$\pm
11$
 \\
 \hline
$100 - 120$ &  0.00162 &$\pm$6.8&$\pm 3.0$ & 0.00035
&$\pm$78&$\pm$ 4
 \\
 \hline
$120 - 140$ &  0.000959 &$\pm$8.0&$\pm 3.8$ &  0.00038
&$\pm$57&$\pm$19
 \\
\hline
\hline
  & \multicolumn{  3}{|c||}{$Q^2> 10\,000$ GeV$^2$} & \multicolumn{ 3}{c}{}\\\cline{1-4}
 $10 - 40$&   0.0099
&$\pm$52&$\pm 6$ &\multicolumn{ 3}{c}{}\\\cline{1-4} $40 - 60$&
0.0017 &$\pm$100& $\pm 4$ &\multicolumn{ 3}{c}{}\\\cline{1-4} $60
- 80$ & 0.0038 &$\pm$89&$\pm$6 &\multicolumn{ 3}{c}{}\\\cline{1-4}
$80 - 100$ & $0$ &\multicolumn{2}{|c||}{$<0.002~(68\%~\textrm{CL})$}&
\multicolumn{ 3}{c}{}\\\cline{1-4}
 \end{tabular}
 \end{center}
\caption{Normalized dijet event cross sections as a function of
$\theta_{\textrm{fwd}}$ and $\theta_{\textrm{bwd}}$. The selection criteria are
given in table 2. The relative statistical errors $\delta_{stat}$
and systematic errors $\delta_{sys}$ are given in per cent.}
\label{tab.th}
 \end{table}



\clearpage\newpage

 \begin{table}[h!]
 \begin{center}
 \begin{tabular}{|c||c|r|r||c|r|r|  }
 \hline
 $z_{p}$
&$\frac{\dps 1}{\dps \sigma_{DIS}}\frac{\dps {\rm
d}\sigma_{2}}{\dps {\rm d}z_{p}}$
&$\delta_{stat}$(\%)&$\delta_{sys}$(\%) &$\frac{\dps 1}{\dps
\sigma_{DIS}}\frac{\dps {\rm d}\sigma_{2}}{\dps {\rm d}z_{p}}$
&$\delta_{stat}$(\%)&$\delta_{sys}$(\%)
\\ \hline
 & \multicolumn{  3}{|c||}{$150<Q^2<275$ GeV$^2$} & \multicolumn{ 3}{|c|}{$275<Q^2<575$ GeV$^2$}\\
 \hline
$0 - 0.125$&  0.377 &$\pm$3.0&$\pm 5.4$ & 0.527 &$\pm$3.4&$\pm
6.1$
 \\
 \hline
$0.125 - 0.25$& 0.620 &$\pm$2.5&$\pm 5.8$ &0.728 &$\pm$2.9&$\pm
3.8$
 \\
 \hline
$0.25 - 0.375$ & 0.474 &$\pm$2.8&$\pm 4.4$ & 0.576 &$\pm$3.2&$\pm
2.2$
 \\
 \hline
$0.375 - 0.5$ &  0.412 &$\pm$3.0&$\pm 5.2$ &  0.514 &$\pm$3.4&$\pm
4.0$
 \\
 \hline
 \hline
  & \multicolumn{  3}{|c||}{$575<Q^2<5000$ GeV$^2$} & \multicolumn{ 3}{|c|}{$5000<Q^2<10\,000$ GeV$^2$}\\
 \hline
$0 - 0.125$&   0.681 &$\pm$4.2&$\pm 4.1$ & 0.65 &$\pm$30&$\pm 7$
 \\
 \hline
$0.125 - 0.25$&   0.804 &$\pm$4.0&$\pm 3.0$ & 0.74 &$\pm$29&$\pm
10$
 \\
 \hline
$0.25 - 0.375$ & 0.710 &$\pm$4.1&$\pm 7.1$ &1.0 &$\pm$26&$\pm 3$
 \\
 \hline
$0.375 - 0.5$ & 0.685 &$\pm$4.2&$\pm 4.9$ &0.88 &$\pm$27&$\pm 7$
 \\
\hline
\hline
  & \multicolumn{  3}{|c||}{$Q^2> 10\,000$ GeV$^2$} & \multicolumn{ 3}{c}{}\\\cline{1-4}
 $0 - 0.125$& 1.2
&$\pm$63&$\pm 3$ &\multicolumn{ 3}{c}{}\\\cline{1-4} $0.125 -
0.25$&0.52 &$\pm$89& $\pm 6$ &\multicolumn{ 3}{c}{}\\\cline{1-4}
$0.25 - 0.375$ & 0.27 &$\pm$100&$\pm$5 &\multicolumn{
3}{c}{}\\\cline{1-4} $0.375 - 0.5$ &1.1 &$\pm$69&$\pm$16
&\multicolumn{ 3}{c}{}\\
\hline
\hline
\cline{1-7}
%
%
 $x_p$
&$\frac{\dps 1}{\dps \sigma_{DIS}}\frac{\dps {\rm
d}\sigma_{2}}{\dps {\rm d}x_p}$
&$\delta_{stat}$(\%)&$\delta_{sys}$(\%) &$\frac{\dps 1}{\dps
\sigma_{DIS}}\frac{\dps {\rm d}\sigma_{2}}{\dps {\rm d}x_p}$
&$\delta_{stat}$(\%)&$\delta_{sys}$(\%)
\\ \hline
  & \multicolumn{  3}{|c||}{$150<Q^2<275$ GeV$^2$} & \multicolumn{ 3}{|c|}{$275<Q^2<575$ GeV$^2$}\\
 \hline
$0 - 0.2$& 0.22 &$\pm$3&$\pm 10$ & 0.12 &$\pm$6&$\pm 13$
 \\
 \hline
$0.2 - 0.4$& 0.368 &$\pm$2.7&$\pm 8.6$ & 0.338 &$\pm$3.5&$\pm 8.2$
 \\
 \hline
$0.4 - 0.6$ &  0.317 &$\pm$2.7&$\pm 7.3$ & 0.424 &$\pm$3.2&$\pm
5.5$
 \\
 \hline
$0.6 - 0.8$ &  0.23 &$\pm$3&$\pm 12$ &0.415 &$\pm$3.0&$\pm 6.3$
 \\
 \hline
$0.8 - 1.0$ &  0.04 &$\pm$5&$\pm 12$ & 0.17 &$\pm$4&$\pm 10$
 \\
 \hline
 \hline
  & \multicolumn{  3}{|c||}{$575<Q^2<5000$ GeV$^2$} & \multicolumn{ 3}{|c|}{$5000<Q^2<10\,000$ GeV$^2$}\\
 \hline
$0 - 0.2$&  0.039 &$\pm$13&$\pm 22$ & 
\multicolumn{3}{c}{}
 \\
 \hline
$0.2 - 0.4$&0.190 &$\pm$6.2& $\pm 9.5$ &0 &
\multicolumn{2}{|c|}{$<0.03~(68\%~\textrm{CL})$}
 \\
 \hline
$0.4 - 0.6$ & 0.348 &$\pm$4.8&$\pm 5.1$ & 0.11 &$\pm$52&$\pm 29$
 \\
 \hline
$0.6 - 0.8$ & 0.573 &$\pm$3.8&$\pm 6.4$ & 0.25 &$\pm$37&$\pm 14$
 \\
 \hline
$0.8 - 1.0$ &  0.65 &$\pm$3&$\pm 10$ & 1.7 &$\pm$18&$\pm$ 4
 \\
 \hline
 \hline
  & \multicolumn{  3}{|c||}{$Q^2> 10\,000$ GeV$^2$} & \multicolumn{ 3}{c}{}\\\cline{1-4}
$0.6 - 0.8$ & 0 &
\multicolumn{2}{|c||}{$<0.2~(68\%~\textrm{CL})$}
&\multicolumn{ 3}{c}{}\\\cline{1-4} $0.8 - 1.0$ &  2.0 &$\pm
47$&$\pm 46$ &\multicolumn{ 3}{c}{}\\\cline{1-4}
 \end{tabular}
 \end{center}
\caption{Normalized dijet event cross sections as a function of
$z_p$ and $x_p$, determined with the modified Durham algorithm.
The selection criteria are given in table 2. The relative
statistical errors $\delta_{stat}$ and systematic errors
$\delta_{sys}$ are given in per cent.} \label{tab.zpxp}
 \end{table}





\clearpage
\newpage

\clearpage\newpage 

\end{document}